# Computational Estimate Visualisation and Evaluation of Agent Classified Rules Learning System

Kennedy E. Ehimwenma, Martin Beer and Paul Crowther
Sheffield Hallam University, United Kingdom

*Abstract*—**Student modelling and agent classified rules learning as applied in the development of the intelligent Pre-assessment System has been presented in [10],[11]. In this paper, we now demystify the theory behind the development of the pre-assessment system followed by some computational experimentation and graph visualisation of the agent classified rules learning algorithm in the estimation and prediction of classified rules. In addition, we present some preliminary results of the pre-assessment system evaluation. From the results, it is gathered that the system has performed according to its design specification.**

*Index Terms*—**agent learning, speech acts, ontology, classification, pre-assessment, student evaluation, visualisation, prediction, artificial intelligence**

## I. INTRODUCTION

Learning is change in the mental state of humans or machines after a sequence of some acquired experiences. Whether these experiences has caused any changes in the "*knower*" is left to be determined by some form of assessment. Learning can be permanent or temporary — meaning that a concept or process can be learned or unlearned. One way to determine the occurrence of learning is through some form of assessment in order to ascertain whether a concept is learned or has been unlearned.

Like humans, machines have the ability to learn. But these abilities are inherent in the chosen type of learning technique. For machines to learn, models—mathematical or symbolic—are chosen or developed suitably to match or solve a learning problem. In this work we have used classification learning in a multiagent system (MAS) for pre-assessing and predicting students' true state of cognition for appropriate leaning materials based on some measurable modelled parameters. The act of using existing knowledge, features or trained examples to make decision is classification learning. Aside having predefined knowledge (or beliefs) for decision making, an agent acquires new knowledge either from self-perception of activities in its environment or through peer-to-peer communication by speech act performatives [1], [22] within a multiagent system. Both predefined knowledge and acquired knowledge amounts to a rise in agent knowledge base (KB) or belief base (BB).

In this paper, we now present in details the theory behind the Pre-assessment System design, the principles applied in the development of the classified rules as well as some computational experimentation and graph visualisation of the agent classified rule learning algorithm and how they make accurate prediction for the required number of classified rules. Also we present the preliminary results of the pre-assessment system evaluation in which the results showed that the system has performed according to its design specification. As revealed from this experimentation, the learning algorithms only holds for a regular ontology i.e. an ontology with equal number of leave-nodes across all parent class nodes [11].

The hallmark of this work is the use of description logic tool – Jason AgentSpeak – in the development of an intelligent tutoring system (ITS) in which agents communicate interoperable knowledge in the format of triples, thus causing changes in their mental state as they carry out the overall system's objective—which is to identify gaps in human learning.

This paper continues with related works in Section I. Section II is BDI: Belief, Desire and Intention in agents, and agent environment. In Section III we present the Pre-assessment agents, and multiple classifications learning in Section IV. Section V presents report on algorithmic experimentation and the results obtained; and Section VI is conclusions and further work.

### A. Related Work: Learning Systems and Strategies of Development

Works in literature has it that several systems has emerged to support learning, teaching, and assessment (LTA). How these systems operate is perhaps determined by the strategy employed in their development e.g. computer assisted assessment (CAA), computer based testing (CBT), intelligent learning system (ILS), computer assisted learning (CAL), computer adaptive testing (CAT), learning management system (LMS) and web-based learning systems. To assess learning for instance, the CBT employs the strategy of presenting predefined sets of questions, while the CAT dynamically select and present questions depending on students' performance [16]. Though varying needs has influenced the design of different systems, holistically, computers in LTA was borne on the need to use technology to support teaching, improve student performance, provide fast and objective marking, change teaching strategies, personalise student instructions, and support ubiquitous and collaborative learning.

Strategies involving intelligent techniques such as agents, machine learning technique and fuzzy logic approaches are also used in developing computer based learning systems. SimStudent [17] was developed with agent technology and machine learning approach. [7] engaged both multiagent and machine learning technique.

Their system used a two- parameter attributes student model: comprehensive ability (C) & problem solving skill (P). In [18], both machine learning technique and multiagent system approach were combined to develop an intelligent system that provided hints to students on current learning goals and prediction of performances. Also there are some research works that provided opportunities to students to recall their prior knowledge before the start of new learning. [26] proposed an intelligent system of this nature where pedagogical agent are meant to evaluate prior knowledge but based on the selective categorisation of users as: novice, beginner, intermediate or advanced learners. The drawback of this is that users make the decision to select the category that they think best fit-into before the presentation of learning materials. In our opinion, self-categorisation may not reveal the actual knowledge status or capability of the user, as users may misjudge the best learning category that may suit their learning needs. Instead, such classification or categorisation should be done by machine intelligence. [24], [23] in a collaborative team project research with the "Guardian Agent" used given *ground rules* in facilitating students participation in online group tasks. Results obtained in [24] showed that the Guardian Agent supported students to identify the module area in which they are well-skilled and so were allocated to the appropriate project group. However, the areas in which some lack-of-skills were indicated by the students' selection of skilled areas, the Guardian Agent did not address.

From the limitations in the foregoing literature, the task of this work was to develop a pre-assessment system with agents in classification learning to categorise users based on some learned parameters like making learning material prediction either for a passed pre-assessment or for a failed pre-assessment. In this view, the prediction of appropriate learning materials after pre-assessment on some prerequisites would allow students to either proceed to learn their preferred area of desired or skilled concept; or learn materials in the area in which some lack-of-skills were identified. Learning the lack-of-skill concept(s) would enable the students to fill-in the gaps in their knowledge.

## II. BELIEF, DESIRES & INTENTIONS

An agent is a computer system that is situated in some environment, and capable of autonomous action in this environment in order to meet its design objectives [25]. Intelligent agent architectures are modelled to have BDI—Belief, Desires and Intentions. BDI is a model of human behaviour, and Jason AgentSpeak is one of those languages that is based-on and inspired-by the BDI model—the idea that projects computer programs to have a mental state [5].

*Beliefs* represent the information agent has about itself, other agents, and its environment [6], [19]. *Desires* represent the tasks allocated to the agent, this corresponds to the objective or goals the agent should accomplish which in effect causes a change in the future states or beliefs of the agents and their environment [6], [4]. *Intentions* represent desires that the agent is committed to achieving [4]. In Jason, the BDI model is accomplished through program plans—some given courses of actions.

Within their environment, agents engage in communicative action to meet their design purpose where they apply practical reasoning approach: reasoning directed towards action [25]. This approach which are used by Jason agents such as in the Pre-assessment System entails what state of affairs to achieve and how to achieve it through plans so that agents are given their: 1) initial beliefs, 2) goals to achieve their intentions, and 3) updated beliefs from the execution of some given goals or plans.

### A. Multiagent System Communication

For agents in a MAS to fulfil their property of cooperation, they must communicate understandably to achieve their collective goal. In such communication, there exist the:

- sender;
- receiver;
- information content;
- intention [designated by performative e.g. tell, achieve, askOne];
- conventions [i.e. messages, negotiations about their goals and actions];
- agent modelling [e.g. their beliefs, goals, authorities, etc.,] in the organisation or environment that they are part of. [2]

To apply the convention of message exchange, the sender, receiver, content, and the intention of communication must be specified. On the Pre-assessment System, the dynamics of interaction and communication starts from the user who enters a concept to learn, through to all the reactive agents and back to the user after the agents has performed their designed specifications. According to FIPA (Foundation for Intelligent and Physical Agents) standard, such communications must be stated in sequence from agent to agent. In Figure 1, we present a FIPA-Compliant Agent communication Flow diagram. The diagram depicts both the static structure and dynamic interaction of the pre-assessment agents.

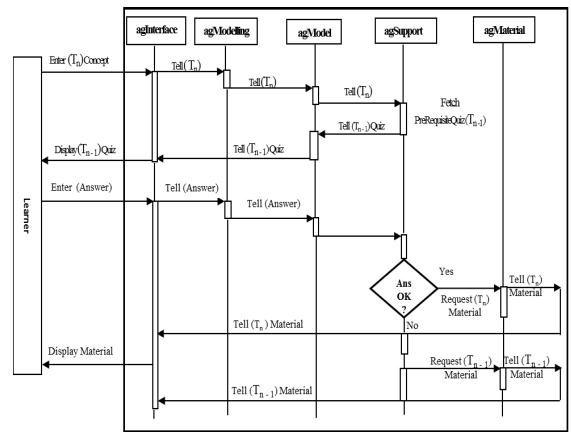

Figure 1. FIPA-Compliant Agent Communication Flow

The diagram showed the detail message passing convention in which performatives are used for communication of knowledge from agent to agent and the decision-making stage of pre-assessment by the agent *agSupport*.

### B. MAS & Environment Programming

One of the properties of agents is that they reside in an environment from where they get percept through sensors (methods in Java), and thereafter act on the percept via actuators (body of a *plan*). To program a multiagent system (MAS), [21] proposed the equation:

*Programming MAS = progr agents + progr environments*

with the view that the two sides of the equation are programs, but with the environment programming part strongly integrated with the agent programming part. Critical examination of the representation made in the equation reveals its conformity with the definition of agent proposed in [25] that — an agent is a computer system that is situated in some environment. That, in that environment they exhibit some properties of autonomy, sociability, cooperation, etc., in order to meet their design objectives. They can observe and perceive the state of the environment that they are situated in, and in effect perform the actions assigned. These environments form a range of artifacts in which agents can render their services.

### C. The Cartago Environment

CArtAgO: *Common Artifact Infrastructure for Agent Open* environment infrastructure are runtime devices providing some kind of function or service in which agents can fruitfully use─both individually and collectively─to achieve their individual as well as social objectives [20]. CArtAgO is a tool for programming and executing artifact based environments: it is a Java based programming model for defining artifacts. To develop the Pre-assessment System user interface, CArtAgO was chosen as the artifact and was configured for the agent *agInterface* to observe users' text-based inputs and interactions.

In some agent systems, CArtAgO has been used to perceive the dynamic changes from iterative mouse clicks precepts [21] of the environment from within an internal-event generating Java program. In the context of this work, we have adopted the approach by customising CArtAgO to perceive dynamic open-ended text-based inputs i.e. SQL queries and concepts which are external (from keyboard devices) to the agents.

### D. Pre, Post & Completion Conditions

The *speech acts* theory of [1] and Searle [22] has predominantly influenced the development of Agent Communication Languages (ACL) such that current speech-act based ACLs specify domain knowledge representation and perfomative communication acts. Labrou & Finin [15] semantics of speech acts shed more light on the *locutionary*, *illocutionary* and *perlocutionary* acts. These [15] described as three performative conditions for agent communication which are represented as *preconditions*, *postconditions* and *completion*:

- **Preconditions**: The fact that is established before an act is performed (i.e. utterance).
- **Postconditions**: The fact that is established after the act is performed (i.e. action).
- **Completion**: The fulfilment of the intention of the act performed (i.e. effect).

#### 1) Establishment of Goals from Speech Acts Paradigm:

Following the [15] semantics of speech acts, three performative conditions for agent communication of goals were established for the Pre-assessment System in preconditions, postconditions and completion. The *completion* Condition becomes the *transition state* [3] in which the agents of the Pre-assessment System can establish the eventual goal that can only be achieved at some time after any current conversation has finished. In the following

*Pre*, *Post* and *Completion* analytics, we present the phases of the semantics of speech acts performatives as they apply to the Pre-assessment System:

i) As a registered student, a student can enter a desired_Concept (to learn) without a precondition, and the set of Preconditions, Postconditions and Completion are as follows:

*Precondition*: student $<L_n>$ has decided on the desired_concept $<T_n>$ to learn.
*Postcondition*: concept $<T_n>$ has been entered.
*Completion*: concept $<T_n>$ has been sent.

ii) But we do not know if the student has adequate prerequisite knowledge to the concept entered to learn. The agent *agSupport* received the concept, and triggers the appropriate plan to sort out the quiz of the prerequisite $<T_{n-1}>$ to the concept $<T_n>$:

*Precondition*: agent agSupport has the rule to sort out the prerequisite quiz $<T_{n-1}>$.
*Postcondition*: agent agSupport sort quiz $<T_{n-1}>$.
*Completion*: quiz $<T_{n-1}>$ is sent for the student's pre-assessment.

iii) The pre-assessment quiz is presented to the student:

*Precondition*: quiz $< T_{n-1}>$ concept has been asked.
*Postcondition*: student has provided an answer $<A_n>$.
*Completion*: KB records updated.

iv) The student's response is communicated back to the agSupport agent:

*Precondition*: student has given a response.
*Postcondition*: agent agSupport tested if answer $<A_n>$ is OK.
*Completion*: student committed to learn.

v) agent agSupport feedback the result of assessment to student:

*Precondition*: student has passed or failed.
*Postcondition*: feedback has been given.
*Completion*: KB Records updated.

vi) agent agModelling has classification attributes

*Precondition*: student attributes received.
*Postcondition*: student has been classified.
*Completion*: classified ontology information is sent to agMaterial.

vii) Appropriate learning to be recommended.
For a Passed result:

*Precondition*: student has *Passed* the quizzes.
*Postcondition*: student is prepared to learn desied_Concept $<T_n>$.
*Completion*: student gets desired_Concept $<T_n>$ URL.

For a Failed response:

*Precondition*: student has *NOT Passed* the quizzes.
*Postcondition*: student is NOT prepared to learn desired_concept $<T_n>$.
*Completion*: student gets prerequisite $<T_{n-1}>$ URL.

This is the semantic analysis of the pre-assessment process.

### E. A Regular Ontology

Ontology is a process of knowledge representation that helps to visualise domain knowledge, its associated con-cepts and the relationships that exist between the concepts. The essence of ontology is to specify true and valid relations or properties that exists between objects in a logical ideology [8]. [12], [13] states that ontology specifies the classes of objects that exist, the relationships amongst those classes, the possible relationships amongst instances of the classes, and constraints over those instances.

This work encompasses agents' use of a *regular ontology:* Ontology with equal number of leaf-nodes across all parent nodes [11]. The ontology is that of a learning structure constructed in the domain of SQL with the Protégé 4.3 OWL ontology editor [14]. Protégé is an ontology construction tool for the semantic web. In the SQL ontology, concepts are interlinked by means of Object Property and Data Property relations, respectively. Notably, the Object Property relation (i.e. *hasPrerequisite*) was used as the predicate relation to form the prerequisite interdependence of a lower-concept in the hierarchy of class structure to its immediate higher-level concepts, and the Data Property relation (i.e. *hasContent*) as the predicate relation for assigning web URL data values to subclass instances or leaf-nodes. The ontology construction consequently lead to the modelling and initialisation of the ontology structure in the BB of the agent agMaterial using the *tagname* of concepts (e.g. *delete*), instead of fully qualified OWL URI (universal resource identifier) names e.g.

*<http://www.sql.com/ontologies/sql.owl#delete>.*

This URI is the namespace of the *delete* concept from Sesame OpenRDF Workbench Repository after the SQL ontology upload (Fig. 2). Before this upload to Sesame, the ontology has been constructed with Protégé.

## III. THE PRE-ASSESSMENT AGENTS

### A. Agent agInterface

This is the agent that is given the *focus* to observe the dynamic user inputs at the artifact CArtAgO. An example of the SELECT input *perception* process is:

```
+value(V)[source(percept)] : value("SELECT")
<-.println("The topic you have entered to learn is: ",
V);
.broadcast(tell, value(V)).
```

### B. Agent agModelling

This is referred to as the *classifier*. It learns and classifies the attributes received from the agent agSupport, and in-turn communicate the agent agMaterial after classification.

### C. Agent Student Model

This is the agent that constructs and keep track of every student activity that is received from the agent *agSupport*. The Student Model agent is configured with the *Jason TextPersistentBB class.* The TextPersistentBB is a persistent text file that captures all activities or learning history which consists of students' desired concept, pre-assessment questions, and correct and/or incorrect answers to questions. These parameter information are also *Time* and *Date stamped* from the agent *agSupport* so that the course tutor can deduce the amount of time a given student has spent on each task.

To identify gaps in students' learning, we have devised a Student Model to keep four parameter-information persis-

Figure 2. A snapshot of fully qualified OWL URIs from the Sesame Ontology Workbench Repository.

tently about a given student. In a tuple, this model has been presented as: M = <D, P, F, V> [10], [11] where

M: is the model
D: a set of desired concepts i.e. desired state
P: a set of passed pre-assessment i.e. current state gains
F: a set of failed pre-assessment i.e. current state gaps
V: the set of SQL query statements.

Parameters <D, P, F> are simultaneously communicated by the agent *agSupport* to the agents *Student* and *agModelling*. The parameterised information are then gathered, learned by the agent agModelling as *pre-conditions* within which the appropriate plan is selected to classify students and make prediction for their learning materials. The rule in a Jason plan format for this classification is given below and some exemplary code in Section IV:

*+recommend_material : set_of_profile_parameters*
*<- recommended_material.*

### D. Agent agSupport

This is the *teacher* in terms of machine learning. It pre-assesses the student based on the desired_concepts received and communicate the outcome of assessments to the agents *agModelling* and *agModel*, respectively. This agent also connects the MAS to MySQL database engine for result-set queries through the *JDBC PersistentBB Driver.* Thus far, from users' queries, the agSupport can make changes to the *Tennis_Database* tables with correct INSERT query statements are logged-in, and display of result-set queries from SELECT query statements. This agent also *Time* and *Date stamped* the outcome of pre-assessments before passing the information to other agents. It also asks the ontology agent, whether the concept it received exist in its BB. An exemplary code is:

```
//plan to receive the SELECT concept. SELECT has no
prerequisite

+value(V)[source(agInterface)] : value(V)==
value("SELECT")                              <-
.date(YY, MM, DD); .time(HH, NN, SS);
.send(agModelling, tell, desired_Concept(V));
.send(student, tell, desired_Concept(V)); .concat(V, ",
date(",YY,"-", MM,"-", DD, ")", ", ", "time(",HH, "-",
NN, "-", SS, ")", Ms); .send(student, tell,
desired_Concept(Ms)); .send(agMaterial, askOne,
hasPrerequisite(V, select));//Asking whether concept
exists .println(V, " has No prerequisite.");
.send(agModelling, tell, recommendMaterial).
```

### E. Agent agMaterial

This is the *ontology* agent that has all ontological relations initialised in its BB including the web URL data value of all SQL concept in the ontology tree. This agent learns its ontological relations in its BB and outputs the appropriate URL learning material after communication from the classifier—agent agModelling. Its other function is to match a users' desired concepts with its BB ontology facts in order to ascertain whether that concept exist, and in-turn inform the user.

## IV. MULTIPLE CLASSIFICATIONS LEARNING

As the *classifier*, the agent *agModelling* learns every attribute of the parameters received from the agent *agSupport* during the course of pre-assessment (Fig. 3). The classification and learning process uses the parameterized attributes described in Section III. Below we give some exemplary classification code in Jason from the agent *agModelling* plan library. This would pre-assess students on the INSERT prerequisite when DELETE is received as the *desired_Concept* [10], [11]:

```
...
/* Prediction rules for DELETE concept */
@d1
+!recommendMaterial[source(agSupport)] : de-
sired_Concept("DELETE")[source(agSupport)]
      & passed("The student has passed the
      INSERT with SELECT question.")
      & passed("The student has passed the
      INSERT with VALUE question.")
<- .send(agMaterial, achieve,        hasPrereq-
uisite(delete, insert)).

@d2
+!recommendMaterial[source(agSupport)] : de-
sired_Concept("DELETE")[source(agSupport)]
      & passed("The student has passed the
      INSERT with SELECT question.")
      & failed("The student has NOT passed the
      INSERT with VALUE question.")
<- .send(agMaterial, achieve, has_KB(insert,
      insert_value)).

@d3
+!recommendMaterial[source(agSupport)] : de-
sired_Concept("DELETE")[source(agSupport)]
      & failed("The student has NOT passed the
      INSERT with SELECT question.")
      & passed("The student has passed the
      INSERT with VALUE question.")
<-.send(agMaterial, achieve, has_KB(insert, in-
sert_select)).

@d4
+!recommendMaterial[source(agSupport)] : de-
sired_Concept("DELETE")[source(agSupport)]
      & failed("The student has NOT passed the
      INSERT with SELECT question.")
      & failed("The student has NOT passed the
      INSERT with VALUE question.")
<-.send(agMaterial, achieve, hasPrerequi-
site(insert, select)).
...
```

In Figure 3, we replicate the mechanism of the code snippet (above) and also show the decision components of two other agents in their dynamic and selective decision-processes and communications. In Jason, agents like humans have mental capabilities. Thus, a *tell* or *broadcast* performative type of message content becomes knowledge to an agent until the MAS is stopped. These semantic information or knowledge which are contained in the BB of the agent forms the basis upon which decisions are made when such knowledge are referenced and satisfied from within the relevant *plan*. So for the classifier agent, the number of parameterised attributes (of the student) that forms each plan (containing a group of semantic knowledge) in the array of classified rules is dependent on the number of leaf-nodes in an ontology structure.

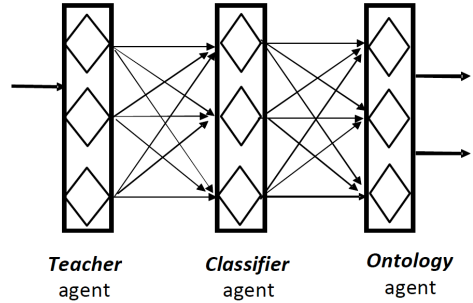

Figure 3. *One vs. All* Multiple Classification

Furthermore, on the code snippet above, students are classified for learning material into one of four categories for the given *desired_Concept* DELETE (in this case) after pre-assessment on its prerequisites. In the code, the attributes of the students which forms the production-rules (otherwise known as the *context* in Jason agentSpeak) or pre-conditions must be true and satisfied before classification is completed.

### A. Agent Classified Rule Learning Estimation Algorithm

[10] states that: In production rules classification learning, let $C$ be the number of prerequisite concept(s) to a desired concept $D$, $T$ a binary-state value for student pre-assessment outcome and $N$ the equal number of leaf-nodes across each parent node, then the total number of classified production rules $R$ *(initialisation)* equation for a given ontology tree is determined by:

$$R = CT^N + 1 \ldots \text{eq.1}$$

where
$$C \in \{0, 1, 2, ..., k\}$$
$$T = 2, \text{ for a pass or fail state}$$
$$N \in \{1, 2, 3, ..., k\}$$

For any SQL rules set that would need to be added to the array of classified rules, the agent *agModelling* would *increment* the number of classified rules for a given concept by:

$$R' = R + CT^{(N-1)} \ldots \text{eq.2}$$

where

$$C = 0, 1, 2, ..., k$$

in $R = CT^N + 1$;

and conversely decrements by removing rules for a concept that is no longer needed with:

$$R' = R - \lceil CT^N / 2 \rceil \ldots \text{eq.3}$$

*where*
$$C \neq 0$$

in $R = CT^N + 1$.

Here it is pertinent to note that (2) and (3) has been slightly modified from those that were first presented as Agent Learning Hypothesis in [10] due to complexities experienced in the accurate scaling of the algorithm; details explained in Section V of this paper.

From each learning algorithm, the number of rules to be added or removed is determined by the number of leaf-nodes $T^N$ in the ontology. Since $T^N = 2^2$ then the number of classified rules equals 4 for each parent class concept of ontology of two-leave nodes. In the DELETE example (Section IV), the agent *agModelling* classifies the student and make prediction for appropriate learning URL through semantic literal communication to the agent *agMaterial* using the *tell* or *achieve* performative. That is,

```
.send(agMaterial, achieve, hasPrerequi-
site(delete, insert));
```

in which the agent agModelling is sending an achieve performative message to the agent agMaterial. The achieve message is a command going by [1] and [22] illucationary acts. On receiving this message, the receiver agent agMaterial execute the action. The achieve message does not form a belief in the receiver's BB, and that is quite different from the following communication:

```
.send(agModelling, tell, passed(X));
```

where the agent agSupport is informing the agent agModelling – via the tell performative. When the receiver agent gets the message, the message becomes a belief thus adding to the agent experience for influencing its classification learning.

### B. Algorithmic Scalability

In the context of this work, an algorithm is said to be scalable if it is suitably efficient and practical when applied to a large number of class node in an ontology, and would estimate the accurate number of classified production rules that is required by an agent to predict and make accurate classification. If the design of an algorithmic system or model fails when some quantity increases then it does not scale. To test for scalability, we chose the graph plotting tool of Python27 programming language.

To describe the scalable element of the algorithms, let's restate the expression for $R$ in (1) above as $R(C, N)$, which implies that $R$ is a function of $C$ and $N$, that is:

$$R(C, N) = CT^N + 1 \ \dots \ \text{eq.4}$$

where

$$T = 2$$

for a two-state value constant of *pass* or *fail*. Then substituting for $T$, (4) becomes:

$$R(C, N) = C * 2^N + 1 \dots \text{eq.5}$$

Thus, its scalability with respect to the element $C$ that is subject to incremental changes, we state

$$R(C', N) = C' * 2^N + 1 \ \dots \ \text{eq.6}$$

where

$C'$ is the prerequisite class nodes as well as the scalable element.

The effect of this is that for sequential increases in the number of class nodes in a regular ontology, the total number of classified rules $R$ for the Classifier agent will equate to $R(C', N)$ i.e. as $C$ increases, $R(C', N)$ estimates the accurate number of classified rules needed by the classifier agent for the accurate classification of the users of the system. Thus, $R(C, N)$ is dependent on $C'$ and $N$; where $N \neq 0$ and must be kept constant and equal across all parent class nodes $C$. See illustrations in Figure 4.

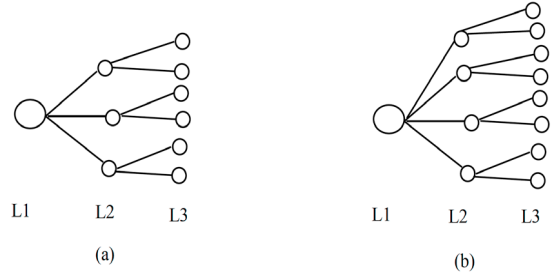

Figure 4.  Depicts scalability of class or parent nodes at the level L2 from (a) to (b).

For scalability with respect to increased changes to $N$, we state that

$$R(C, N') = C * 2^N + 1 \dots \text{eq.7}$$

such that $R(C, N')$ determines the accurate number of classified rules while $C$ is kept constant. This is illustrated in Figure 5.

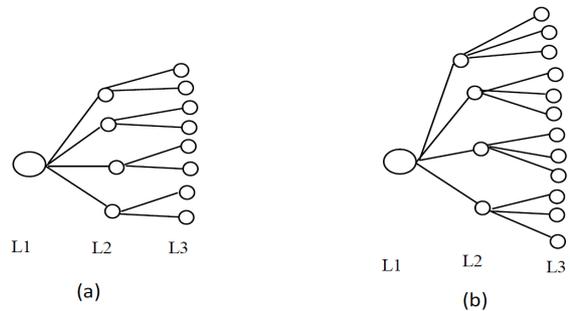

Figure 5.  Scalability of leave-nodes N at level L3 of (a) to (b)

### C. Principle of Classified Rule formulation

Considering Figure 4(a & b), we have three levels L1, L2 and L3 in the ontology; and that is also true of the Figure 5(a & b). In Figure 4(a), we say the ontology tree has three parent class nodes but two prerequisites classes $C$ at level L2, and in Figures 4(b), 5(a) and 5(b) at level L2, the ontology trees has four parent class nodes but three pre-requisite concepts $C$, respectively. Any of the parent node at Levels L2, can be a desired_Concept $D$ and the nodes beneath them their prerequisite(s).

Now considering the <P> and <F> parameters of the *Student* model tuple, for a desired_Concept that has no prerequisite i.e. C = 0, the total number of rules R = 1 (the DEFAULT). This is the default rule that would output only the web URL link for that concept when the least concept is the *desired_Concept* of the student. But for other higher concepts, we now describe the agent classified rules formulation process as follows:

For a Desired_Concept $D$ with:

a) prerequisite $C = 1$, leaf-node $N = 1$, the number of rules $R = 3$.
That is:

$$C1 = \frac{P}{F + 1 \text{ (DEFAULT)}}$$

b) prerequisite $C = 1$, leaf-node $N = 2$, the number of rules $R = 5$
That is:

$$C1 = \begin{matrix} PP \\ PF \\ FP \\ FF + 1 \text{ (DEFAULT)} \end{matrix}$$

c) prerequisite $C = 1$, leaf-node $N = 3$, the number of rules $R = 9$
That is:

$$C1 = \begin{matrix} PPP \\ PPF \\ PFF \\ FPP \\ FFP \\ PFP \\ FPF \\ FFF + 1 \text{ (DEFAULT)} \end{matrix}$$

d) prerequisite $C = 2$, leaf-node $N = 2$, the number of rules $R = 9$
That is:

$$C2 = \begin{matrix} PP \\ PF \\ FP \\ FF \end{matrix}$$

$$C1 = \begin{matrix} PP \\ PF \\ FP \\ FF + 1 \text{ (DEFAULT)} \end{matrix}$$

This process of classified rules formulation is dependent on the $C, T, N,$ and $D$; where $D$ becomes the attribute that will first and foremost prune the search space to the category of the *desired_Concept*.

## V. Experiments & Results

### A. Test-Running Algorithmic Scalability

Having used Python27 as the algorithmic testing and data generating tool, we now present the results from the test of the three equations: *Initialisation*, *Incremental* and *Decremental* equations. The data sets obtained showed accurate predictions in the number of classified rules $R$ needed in an agent's BB. This test was run on a number of iterations and results were manually compared for a range of $N = 1$ to $5$, and for $C = 0$ to $6$. Figure 6 (for example) shows one of the iterative processes and results of how data was generated for the equation

$$R(C', N) = C'^* 2^N + 1$$

as $C$ increases. The result shows that the equation

$$R'(C, N) = C'^* 2^N + 1$$

scales accurately well in predicting the number of rules $R$.

Similarly, accurate results were also obtained when

$$R(C, N') = C^* 2^N + 1$$

in (7) was test-run with sequential increases in $N$. Figure 6 − 8 shows some program execution.

Figure 6.    Result of $R(C, N) = C^* 2^N + 1$

### B. Algorithmic Complexity & Solution

Subject to to extensive testing, we however found some complexity in the equation $R' = R + 2^N$ that was initially given as the *Incremental* algorithm in [10] due to inability to scale accurately to predict the needed number of rules R for all $C$ and $N$. This complexity lead to the re-examination and modification of the *incremental* algorithm given in (2), and the its inverse *decremental* algorithm in (3) above, respectively.

This is because $R' = R + 2^N$ only scaled accurately for the prerequisite $C = 2$ and its sequential increases in $N$, but failed to scale accurately for $C = 0, 1, 3, 4, 5$, etc., when it was computed programmatically (Fig. 7).

Figure 7.    Result of $R' = R + 2^N$ where $C = 2$

The modified scalable equations are thus

$$R'(C, N') = R + CT^{(N-1)}$$

the *incremental* algorithm in (2), and it inverse algorithm---the *decremental* in (3)

$$R'(C, N') = R - [CT^N / 2].$$

The decremental algorithm decrements or reduces the exact rule number $R$ that is given to the point where $N = 1$ accurately. The reduction process is determined by expression

$$CT^N / 2$$

in the equation. The number of rules $R$ left after each decremental computation has yielded accurate number of classified rules $R$ for every $C$ and $N$ (see Fig. 8b). The incremental and reduction process of rules would be induced subject to constraints to be given to agents. In this ontology, no blank node are allowed, because the algorithm will not scale accurately. As a prove, see figure 8(b) where N iterated to zero, the value of $R$ was 5, which in reality should be zero. Thus, $N$ can never take a zero value.

```
Computation of Agent_Classified_Rules Learning Algorithm: R = R + C*(T**(N-1))

A program to estimate the number of rules in an ontology of STRUCTURED-PREREQUISITES but increas

Initial number of rules for the ontology of  1 LEAVE_NODE N with PREREQUISITE C = 3 is:  7
Total number of rules for the ontology of  2 LEAVE_NODE N with PREREQUISITE C = 3 is:  13
Total number of rules for the ontology of  3 LEAVE_NODE N with PREREQUISITE C = 3 is:  25
Total number of rules for the ontology of  4 LEAVE_NODE N with PREREQUISITE C = 3 is:  49
Total number of rules for the ontology of  5 LEAVE_NODE N with PREREQUISITE C = 3 is:  97
Total number of rules for the ontology of  6 LEAVE_NODE N with PREREQUISITE C = 3 is:  193
Total number of rules for the ontology of  7 LEAVE_NODE N with PREREQUISITE C = 3 is:  385
Total number of rules for the ontology of  8 LEAVE_NODE N with PREREQUISITE C = 3 is:  769
Total number of rules for the ontology of  9 LEAVE_NODE N with PREREQUISITE C = 3 is:  1537
Total number of rules for the ontology of 10 LEAVE_NODE N with PREREQUISITE C = 3 is:  3073
Total number of rules for the ontology of 11 LEAVE_NODE N with PREREQUISITE C = 3 is:  6145
Total number of rules for the ontology of 12 LEAVE_NODE N with PREREQUISITE C = 3 is:  12289
Total number of rules for the ontology of 13 LEAVE_NODE N with PREREQUISITE C = 3 is:  24577
Total number of rules for the ontology of 14 LEAVE_NODE N with PREREQUISITE C = 3 is:  49153
Total number of rules for the ontology of 15 LEAVE_NODE N with PREREQUISITE C = 3 is:  98305
Total number of rules for the ontology of 16 LEAVE_NODE N with PREREQUISITE C = 3 is:  196609
end
```

(a) Result for R' = R + C*T**(N − 1); C = 3, N = N + 1

```
Computation of Agent Decremental Learning Algorithm: R = R - C*T**N/2

Estimating the number of rules in an ontology of STRUCTURED-PREREQUISITES by decreasing number of Leave_Nodes.

Total number of rules for ontology with LEAVE_NODE N = 5  and PREREQUISITE C = 4 is:  129
Decremental estimate of number of R by decrease in LEAVE_NODE from  5 to 4 with PREREQUISITE C = 4  is:  65
leave_node N is  4
Decremental estimate of number of R by decrease in LEAVE_NODE from  4 to 3 with PREREQUISITE C = 4  is:  33
leave_node N is  3
Decremental estimate of number of R by decrease in LEAVE_NODE from  3 to 2 with PREREQUISITE C = 4  is:  17
leave_node N is  2
Decremental estimate of number of R by decrease in LEAVE_NODE from  2 to 1 with PREREQUISITE C = 4  is:  9
leave_node N is  1
Decremental estimate of number of R by decrease in LEAVE_NODE from  1 to 0 with PREREQUISITE C = 4  is:  5
leave_node N is  0
end
```

(b) Result for R' = R − [C*T**N / 2]; C = 4, N = N - 1

Figure 8.

## C. Data Set and Data Visualisation

From the computations, we obtained the following exampler data set for $R$ based on different values of $C$ and $N$, respectively. The data generated are presented in a group of three for $C = 0$ to $6$, $N = 1$ to $5$, and $R = 0$ to $6$ vectors. The values were plotted, and from the plots the behaviour of the algorithms were visualised (Fig. 9 & 10).

C0 = [0, 0, 0, 0, 0]
N  = [1, 2, 3, 4, 5]
R0 = [1, 1, 1, 1, 1]

C1 = [1, 1, 1, 1, 1]
N  = [1, 2, 3, 4, 5]
R1 = [3, 5, 9, 17, 33]

C2 = [2, 2, 2, 2, 2]
N  = [1, 2, 3, 4, 5]
R2 = [5, 9, 17, 33, 65]

C3 = [3, 3, 3, 3, 3]
N  = [1, 2, 3, 4, 5]
R3 = [7, 13, 25, 49, 97]

C4 = [4, 4, 4, 4, 4]
N  = [1, 2, 3, 4, 5]
R4 = [9, 17, 33, 65, 129]
C5 = [5, 5, 5 5, 5]
N  = [1, 2, 3, 4, 5]
R5 = [11, 21, 41, 81, 161]

C6 = [6, 6, 6, 6, 6]
N  = [1, 2, 3, 4, 5]
R6 = [13, 25, 49, 97, 193]

In the data set, it is observed that the values of $R$ has a regular pattern. This is assumed to be connected to the regular ontology structure—representation of equal leaf-nodes. In addition, it is also noticed that the set of data for $R$ follows a progressive trend for $C$ and $N$ respectively. This spontaneously generated an interest in the research to unravel the factors behind the these patterns. Subsequently the need arose to develop the equation(s) to fit this pattern: Equations that can predict the values of $R$.

## D. Preliminary Evaluation of the Multiagent Based Pre-assessment System

In this section, we present the results of the Pre-assessment System evaluation. This evaluation is the preliminary test of the system: to see the performance of the system, check fitness for purpose as well as get users' feedback so as to improve design. Participants in this evaluation were MSc and BSc final year undergraduate Database students respectively. They were recruited for this purpose after giving their consent.

Following the model of design, all student activity and history of learning were persistently recorded in the agent *student* TextPersistentBB beliefs. This BB history was accessed by the researcher and results were analysed. The analysis of the data collated from the BB showed that the system identified learning gaps with respect to the Desired_Concept of students (Fig. 11(a) and (b)).

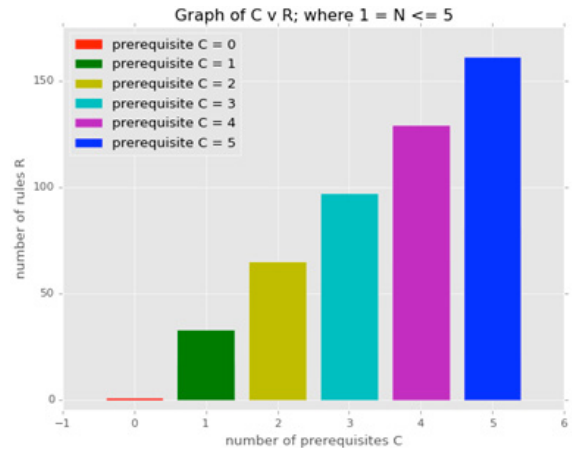

Figure 9. Visualised estimation of R in graph C vs. R. N increases from 1 to 5 for each Prerequisite C class node.

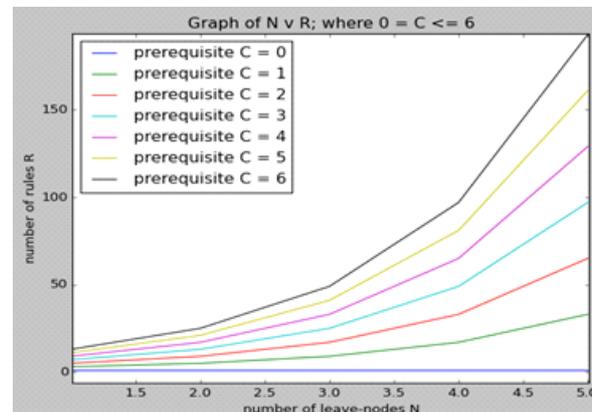

Figure 10. Visualised estimation and behaviour of the graph N vs. R

| Desired_Concept | Questions (Q) | Time Question was asked (HH-MM-SS) | Time Answer was entered (HH-MM-SS) | Outcome of Pre-Assessment | Time spent task (HH-MM-SS) | Remark & Recommendation |
|---|---|---|---|---|---|---|
| | Q1: DELETE_SELECT | | | | | |
| | 1st attempt: | 11-8-54 | 11-9-27 | Not Passed | 00-0-33 | |
| UPDATE | 2nd attempt: | 11-11-31 | 11-12-10 | Not Passed | 00-0-39 | CONCLUSION Not prepared to learn UPDATE, instead was recommended to learn DELETE_SELECT; and DELETE_WHERE. |
| | Q2: DELETE_WHERE | | | | | |
| | 1st attempt: | 11-9-27 | 11-12-10 | Not Passed | 00-2-33 | |
| | 2nd attempt: | 11.12.10 | 11-14-14 | Not Passed | 00-2-24 | |

(a): Student A: Analysis of pre-assessment when UPDATE was entered as Desired_Concept.

| Desired_Concept | Questions (Q) | Time of Question Presentation (HH-MM-SS) | Time of Answer Response (HH-MM-SS) | Outcome of Pre-Assessment | Time spent task (HH-MM-SS) | Remark & Recommendation |
|---|---|---|---|---|---|---|
| | Q1: OUTER_JOIN | | | | | |
| | 1st attempt: | 11-28-48 | 11-28-56 | Not Passed | 00-0-08 | |
| UNION | 2nd attempt: | 11-29-48 | 11-31-43 | Not Passed | 00-1-55 | CONCLUSION Not prepared to learn UNION, instead was recommended to learn FULL_OUTER_JOIN; and INNER_JOIN. |
| | Q2: INNER_JOIN | | | | | |
| | 1st attempt: | 11-28-56 | 11-29-35 | Not Passed | 00-0-39 | |
| | 2nd attempt: | 11-31-43 | 11-34-04 | Not Passed | 00-2-21 | |

(b): Student B: Analysis of pre-assessment when UNION was entered as Desired_Concept.

Figure 11.

In Figure 11(a) *Student A* analysis, the UPDATE concept was the desired_Concept. *Student A* was pre-assessed on the DELETE concept which is the prerequisite or immediate lower node to UPDATE. So the pre-assessment started from the DELETE_SELECT to the DELETE_WHERE leaf-node i.e. from the more technical concept to the less technical leaf-node concept. From the analysis *Student A* had two attempts on the same UPDATE desired_Concept. In the first attempt, *Student A* was pre-assessed to have "NOT Passed". Then a second attempt was made, but the same result was obtained; and *Student A* was thereafter recommended to learn both concepts: the DELETE_SELECT and DELETE_WHERE. Thus, the student was not prepared to learn the UPDATE desired_Concept.

Similarly in Figure 11(b) on the analysis of *Student B*, the same scenario played out — to have "NOT Passed" the pre-assessment on the JOIN query leaf-node concepts when *Student B* entered UNION as desired_Concept. *Student B* also had two attempts.

At the end of each pre-assessment exercise, the students were classified and recommended to learn materials of the failed SQL concepts; and the URL links were presented by the ontology agent to fill-in the knowledge gap. Also from the analysis, the time spent on each task by the students were deduced as recorded by the agent *Student*. From Figure 11(a) *Student A* spent an average of *36sec* on the DELETE_SELECT task, and an average of *2min 24sec* on the DELETE_WHERE task. In figure 11(b), on the OUTER_JOIN task, *Student B* spent an average of *1min 01sec*, and an average of *1min 15sec* on the INNER_JOIN. Overall *Student A* spent a total of *2mins* while *Student B* spent a total of *1min 08sec* doing their tasks on the system.

### E. User Feedback from System's Evaluation

After the evaluation of the Pre-assessment system, the students were invited to take part in a survey on Survey-Monkey about their experiences with the system. From the 20 item questionnaire data also collated, it was gathered that the students are quite familiar with SQL, and that the pre-assessment of prerequisite concept(s) when they entered a *Desired_Concept* helped them to recall their SQL skills. However, there was the dissenting view that the system does not give room for "*trial and error*" test on their SQL code. That, it gives no second chance as the system would evaluate your code straight-on as soon as your SQL queries are logged-in; and then to the next question.

## VI. CONCLUSIONS & FURTHER WORK

This paper has demonstrated the use of a multiagent system tool in developing intelligent tutoring and learning system (ITLS) by employing the One vs. All Multiple Classification technique. We have shown how agents learns, reason and share knowledge through semantic communication in order to cooperatively diagnose gap(s) between a student's desired knowledge and his previous knowledge from some devised set of parameters. Our preliminary evaluation of the system showed that the Pre-assessment Systems identified gaps in skills and predicted learning materials. The paper also detailed the development process of the three algorithms that estimates and predicts the required number of classified rules for agents. To the best of our knowledge, these algorithms are the first in a report that estimates classified number of rules based on the use of ontologies. The scalability of the algorithms were tested and shown using graph visualisations. In the future we shall automate the process in which agents can use these algorithms to update their classified rules under some given constraints. We also intend to conduct further studies on multiagent connection, communication and querying of ontology repositories such as Sesame OpenRDF Workbench.

## ACKNOWLEDGEMENT

We acknowledged Jomi Hubner—a co-author of the book "Programming Multi-agent Systems in AgentSpeak using Jason"—for his technical advice on Jason in this work. Federal University of Santa Catarina, Department of Automation and Systems Engineering; Brazil.

## REFERENCES

[1] Austin, J. L. (1962). How to do things with words. Oxford university press.

[2] Barbuceanu, M., & Fox, M. S. (1995). COOL: A language for describing coordination in multi agent systems. Icmas, pp. 17-24.

[3] Bench-Capon, T. J. (1998). Specification and implementation of toulmin dialogue game. Proceedings of JURIX, pp. 5-20.

[4] Bellifemine, F. L., Caire, G., & Greenwood, D. (2007). Developing multi-agent systems with JADE. John Wiley & Sons. West Sussex, England. http://dx.doi.org/10.1002/9780470058411

[5] Bordini, R. H., Hubner, J. F., & Wooldridge, M. (2007). Programming multi-agent systems in AgentSpeak using Jason (Vol. 8). John Wiley & Sons. http://dx.doi.org/10.1002/9780470061848

[6] Bratman, M. (1987). Intention, Plans, and Practical Reason. Harvard University Press: Cambridge, MA.

[7] Chakraborty, S., Roy, D., & Basu, A. (2010). Development of knowledge based intelligent tutoring system. Advanced Knowledge Based Systems: Model, Applications & Research, 1, 74-100.

[8] Ehimwenma, K., Beer, M., & Crowther, P. (2014, April). Ontology Engineering and Modelling for Learning Activity in a Multiagent System. In Proceedings of the 2014 First International Conference on Systems Informatics, Modelling and Simulation (pp. 177-181). IEEE Computer Society.

[9] Ehimwenma, K., Beer, M., & Crowther, P. (2014, July). Pre-assessment and Learning Recommendation Mechanism for a Multi-agent System. In Advanced Learning Technologies (ICALT), 2014 IEEE 14th International Conference on (pp. 122-123). IEEE. http://dx.doi.org/10.1109/icalt.2014.43

[10] Ehimwenma, K. E., Beer, M., & Crowther, P. (2015). Adaptive Multiagent System for Learning Gap Identification Through Semantic Communication and Classified Rules Learning. 7th International Conference on Computer Supported Education. In Doctoral Consortium (CSEDU), pp. 33-38. SCITEPRESS.

[11] Ehimwenma, K. E., Beer, M., & Crowther, P. (2015). Student Modelling and Classification Rules Learning for Educational Resource Prediction in a Multiagent System. 7th Computer Science and Electronic Engineering Conference (CEEC2015), UK.

[12] Gruber, T. R. (1993). A translation approach to portable ontology specifications. Knowledge acquisition, 5(2), 199-220. http://dx.doi.org/10.1006/knac.1993.1008

[13] Gruber, T. R. (1995). Toward principles for the design of ontologies used for knowledge sharing. International journal of human-computer studies, 43(5), 907-928. http://dx.doi.org/10.1006/ijhc.1995.1081

[14] Horridge, M., Knublauch, H., Rector, A., Stevens, R., & Wroe, C. (2004). A Practical Guide To Building OWL Ontologies Using The Protege-OWL Plugin and CO-ODE Tools Edition 1.0. University of Manchester.

[15] Labrou, Y., & Finin, T. (1998). Semantics and conversations for an agent communication language. Readings in Agents, 235-242.

[16] Lilley, M and Barker, T. (2003). An evaluation of a computer adaptive test in a uk university context. In 7th Computer assisted assessment conference, pages 89.

[17] Matsuda, N., Cohen, W. W., Sewall, J., Lacerda, G. &

[18] Koedinger, K. R. (n.d.). SimStudent: Building an Intelligent

[19] Tutoring System by Tutoring a Synthetic Student. Pp.1-19.

[20] http://www.cs.cmu.edu/~wcohen/postscript/simstudent-authoring-submitted.pdf (Accessed: August 20th, 2015).

[21] Mills, C., & Dalgarno, B. (2007). A conceptual model for game-based intelligent tutoring systems. Proceedings of the 2007 Australasian Society for Computers in Learning in Tertiary Education, 692-702.

[22] Padgham, L., & Winikoff, M. (2004). Developing intelligent agent systems: A practical guide John Wiley & Sons. West Sussex, England. http://dx.doi.org/10.1002/0470861223

[23] Ricci, A., Viroli, M., & Omicini, A. (2006). CArtAgO: An infrastructure for engineering computational environments in MAS. Weyns et al.[31].

[24] Ricci, A., Piunti, M., & Viroli, M. (2011). Environment programming in multi-agent systems: an artifact-based perspective. Autonomous Agents and Multi-Agent Systems, 23(2), 158-192. http://dx.doi.org/10.1007/s10458-010-9140-7

[25] Searle, J. R. (1969). Speech acts: An essay in the philosophy of language (Vol. 626). Cambridge university press. http://dx.doi.org/10.1017/cbo9781139173438

[26] Whatley, J.; Beer, M. & Staniford, G. (2000). Facilitation of Online Student Group Projects with a Support Agent. ACM Conference City, State.

[27] Whatley, J. (2004). An Agent System to Support Student Teams Working Online. Journal of Information Technology Education, Volume 3, pp.53-63.

[28] Wooldridge, M. (2009). An Introduction to MultiAgent Systems. John Wiley & Sons Ltd, UK.

[29] Yu, S. and Zhiping, L (2008). Intelligent pedagogical agents for intelligent tutoring systems. In Computer Science and Software Engineering, 2008 International Conference on, volume 1, pages 516519. IEEE. http://dx.doi.org/10.1109/csse.2008.414

## AUTHORS

**Kennedy E. Ehimwenma, Martin Beer** and **Paul Crowther** are with the Communication & Computing Research Centre, Department of Computing, Sheffield Hallam University, United Kingdom (e-mails: Kennedy.K.Ehimwenma@student.shu.ac.uk, M.Beer@shu.ac.uk, P.Crowther@shu.ac.uk).

This article is an extended and modified version of a paper presented at the 5th International Conference on Computer Supported Education (CSEDU 2015), 23-25 May 2015, Lisbon, Portugal. Submitted 29 August 2015. Published as resubmitted by the authors 26 December 2015.